# Resistive Switching and Current Conduction Mechanisms in Hexagonal Boron Nitride Threshold Memristors with Nickel Electrodes

*Lukas Völkel[1], Dennis Braun[1], Melkamu Belete[1], Satender Kataria[1], Thorsten Wahlbrink[2], Ke Ran[3,4], Kevin Kistermann[3], Joachim Mayer[3,4], Stephan Menzel[5], Alwin Daus[1*], Max C. Lemme[1,2*]*

L. Völkel[1], D. Braun[1], M. Belete[1], S. Kataria[1], A. Daus[1], M. C. Lemme[1,2]

[1] Chair of Electronic Devices, RWTH Aachen University, Otto-Blumenthal-Str. 25, 52074 Aachen, Germany.
E-mail: max.lemme@eld.rwth-aachen.de, alwin.daus@eld.rwth-aachen.de

T. Wahlbrink[2], M. C. Lemme[1,2]

[2] AMO GmbH, Advanced Microelectronic Center Aachen, Otto-Blumenthal-Str. 25, 52074 Aachen, Germany.

K. Ran[3,4], K. Kistermann[3], J. Mayer[3,4]

[3] Central Facility for Electron Microscopy (GFE), RWTH Aachen University, Ahornstr. 55, 52074, Aachen, Germany.

K. Ran[3,4], J. Mayer[3,4]

[4] Ernst Ruska-Centre for Microscopy and Spectroscopy with Electrons (ER-C 2), Forschungszentrum Jülich, 52425, Jülich, Germany.

S. Menzel[5]

[5] Peter Gruenberg Institute (PGI-7), Forschungszentrum Juelich GmbH and JARA-FIT, 52425 Juelich, Germany.

Keywords: memristor, threshold switching, hexagonal boron nitride, filament, 2D materials

**Abstract**

The two-dimensional (2D) insulating material hexagonal boron nitride (h-BN) has attracted much attention as the active medium in memristive devices due to its favorable physical properties, among others, a wide bandgap that enables a large switching window. Metal filament formation is frequently suggested for h-BN devices as the resistive switching (RS) mechanism, usually supported by highly specialized methods like conductive atomic force microscopy (C-AFM) or transmission electron microscopy (TEM). Here, we investigate the switching of multilayer hexagonal boron nitride (h-BN) threshold memristors with two nickel (Ni) electrodes through their current conduction mechanisms. Both the high and the low resistance states are analyzed through temperature-dependent current-voltage measurements. We propose the formation and retraction of nickel filaments along boron defects in the h-BN film as the resistive switching mechanism. We corroborate our electrical data with TEM analyses to establish temperature-dependent current-voltage measurements as a valuable tool for the analysis of resistive switching phenomena in memristors made of 2D materials. Our memristors exhibit a wide and tunable current operation range and low stand-by currents, in line with the state of the art in h-BN-based threshold switches, a low cycle-to-cycle variability of 5 %, and a large On/Off ratio of $10^7$.

**1. Introduction**

Recently, hexagonal boron nitride (h-BN) has become a popular material for resistive switching (RS) as it offers favorable properties like high in-plane thermal conductivity, thermal and chemical stability, mechanical flexibility and a wide bandgap (~5.9 eV) that enables a large switching window[1–4] Moreover, the layered van der Waals (vdW) structure with ideally a dangling bond-free surface enables integration on arbitrary substrates and pristine interfaces.[3,5] Previously, 2D h-BN has been reported to exhibit (volatile) threshold switching (TS)[6–8] and non-volatile RS.[3,9–12] In contrast to common metal-oxide RS devices, h-BN typically has crystalline vdW layers that potentially allow controlled device operation through the design of grain boundaries, defects, and layer numbers.[3,6,13] Nevertheless, to the best of our knowledge, many h-BN-based devices reported in literature use materials that are incompatible with conventional silicon technology.[3,7,13–17] Mostly, metal filament formation is suggested for h-BN devices as the RS mechanism,[6,12,13,18,19] usually supported by methods like conductive atomic force microscopy (C-AFM), transmission electron microscopy (TEM) or by analyzing single direct current (DC) current-voltage (*I-V*) curves.[6,13,15–17,20–24] TEM is destructive and often only a small part of a device can be measured. C-AFM probes the RS behavior only locally and does not represent a practical device configuration. A fully fabricated device can only be destructively investigated with C-AFM, where the top electrode must be removed after the switching.[25–27] Electrical measurements, in contrast, can be very useful in understanding the fundamental current conduction mechanisms in memristive devices. They are thus an easily accessible, potentially non-destructive method to extract information about defects, their role in the carrier transport, and insights into the RS mechanism in the entire device.[28] Only a few publications discuss the current conduction mechanism of monolayer h-BN-based devices[9,29] while the current conduction mechanisms in multilayer h-BN memristors are still unknown.

Here, we investigate the current conduction mechanism in Ni/h-BN/Ni cross-point TS devices for both high and low resistance states through temperature-dependent $I$-$V$ measurements. Based on the extracted current conduction mechanism and TEM images, we propose metal-filament formation across the h-BN films and subsequent self-rupture as the TS mechanism in our h-BN memristors. The devices in this study show volatile operating metrics similar to state-of-the-art, including low cycle-to-cycle variability and a large On/Off ratio. They were fabricated with a complementary metal oxide semiconductor (CMOS)-compatible material stack, providing a viable alternative for future integration with CMOS when suitable large-scale transfer methods become available.[30,31]

## 2. Results and discussion

Figure 1a shows a schematic of our fabricated cross-point devices with h-BN sandwiched between two Ni electrodes. The bottom electrode (BE) and the top electrode (TE) were patterned with optical lithography followed by 60 nm Ni sputtering and lift-off. H-BN was transferred using a polymer-assisted wet-chemical transfer method (see Experimental Methods). Two h-BN films were stacked on top of each other through subsequent transfer processes to minimize the formation of cracks and macroscopic defects. Possible photoresist residues or wrinkles from the transfer are not expected to influence the RS behavior in our devices as they only increase the film resistance locally and therefore suppress filament formation only at those locations, as shown for h-BN-based memristive circuits.[32] In particular, the fact that the resist was not exposed to plasma processing prior to its removal minimizes residues. An additional 50 nm thick aluminum (Al) layer was sputtered on top of both Ni contact pads after TE deposition to provide good electrical contact between the device and the probe needles during measurements. The active area of the devices was measured to be 34 μm². A possible impact of nickel-oxide formation between the Ni and the Al at the contact pads on the RS behavior is excluded by *I-V* measurements on Ni/Ni control devices without h-BN between the electrodes (supporting information Figure S2). Figure 1b shows a top-view optical microscope image of a fabricated cross-point device. Reactive ion etching (RIE) was carried out to remove the h-BN films from the Al contact pad area and led to polymer residues. However, this is not expected to influence the device performance as the etching was done after the intrinsic device stack of Ni/h-BN/Ni was fully assembled. More details about the fabrication methods are available in the Experimental Methods section, and a schematic process flow is shown in Figure S1 in the supporting information. Figure 1c presents a top-view optical microscope image of the boundary area showing both the first transferred h-BN film and the stack of the two films. The inset Raman map shows the integrated peak intensity around the characteristic $E_{2g}$ peak of bulk h-BN at 1366 cm$^{-1}$.[33] The region with only one h-BN film is distinguishable from that with two h-BN films due to the overall higher peak intensity for the thicker film. However, the visibly patchy structure indicates a spatial variation of h-BN layer numbers in both regions.[2,33] A histogram of all peak positions from the Raman map and a Gaussian distribution fit are shown in Figure S3 in the supporting information. The mean peak position is at $1367.0 \pm 0.9$ cm$^{-1}$, indicating 3-layer to bulk h-BN.[2] Figure 1d shows four spectra extracted from the positions marked in the Raman map in Figure 1c. The peak position only marginally changes for the different measurement positions, whereas the peak height is increased at the brighter positions compared to the darker ones, confirming a non-uniform h-BN thickness. A scanning electron microscopy (SEM) image of the two stacked h-BN films shows white triangles, identified as multilayer nucleation sites, in front of a darker background, supporting the findings from the Raman measurements (Figure 1e). A high-resolution TEM (HRTEM) cross-section image of a fabricated device is presented in Figure 1f. The layered 2D nature of the multilayer h-BN film is clearly visible with an interlayer distance of $3.3 \pm 0.02$ Å, which is in good agreement with values reported in literature.[34,35] The boundary between the two transferred h-BN films manifests in a deviation of the layer distance between two adjacent layers of almost 30 % (see supporting information section S4). The first and second transferred h-BN films are highlighted in blue and orange, respectively. It was estimated from the HRTEM image that the h-BN devices (consisting of the two transferred films) have an average thickness of $5 \pm 0.5$ nm.

*I-V* characterization was performed on cross-point devices. **Figure 2a** shows five subsequent voltage sweeps from 0 V to 3 V and back to 0 V with a sweep rate of ~0.33 V s$^{-1}$, which were applied to the top electrode of a device while the bottom electrode was grounded (see inset of Figure 2a). The device is initially in a high resistance state (HRS) but changes abruptly to a low resistance state (LRS) during the forward sweep (indicated with arrow number 1) at an on-threshold voltage ($V_{th,on}$) of ~2.1 V. The current compliance (CC) was set to 10 nA for this measurement. The device initially remains in LRS when the voltage is swept back towards 0 V, but eventually switches back to the HRS (arrow number 2), leading to an *I-V* hysteresis. The voltage at which the current level of the backward sweep reaches the initial level of the HRS is defined as the off-threshold voltage $V_{th,off}$. The device exhibited reproducible TS behavior[36] with a current level below the noise limit of the measurement setup (1 pA) for voltages below 0.3 V in the HRS. Such low currents in the HRS open up the possibility to use our devices as selectors for nonvolatile resistive switching cells.[16] As an indirect measure for endurance, we plot the extracted $V_{th,on}$ and $V_{th,off}$ of ~600 subsequent cycles in a histogram (**Figure 2b**). We observe a tight distribution of $V_{th,on}$ and $V_{th,off}$ by fitting a Gaussian distribution curve to the data with the centers at 2 V and 0.3 V, respectively. The distribution width $\sigma$ of 0.1 V for $V_{th,on}$ and $V_{th,off}$ indicates a low cycle-to-cycle variability of $V_{th,on}$ of 5 %, which is in same range of the best prior reported cycle-to-cycle variability of bipolar resistive switching in few-layer h-BN devices.[11] The complete set of *I-V* curves of the endurance measurement is plotted in the supporting information (Figure S5). The inset of Figure 2b shows the normalized cumulative distribution functions (CDF) of the $V_{th,on}$ and $V_{th,off}$ distributions. Our devices exhibit TS behavior over three orders of magnitude of CC (**Figure 2c**). Typical TS behavior is visible for each CC. We extract an average $V_{th,on}$ of 2.0 ± 0.1 V, consistent with the value extracted from the endurance measurement, which implies that the CC does not influence the cycle-to-cycle variability considerably. The off-current of our devices (at a read voltage of 0.1 V) are below the noise limit of our measurements in Figure 2c, where high currents must be resolved. We measured the current level before and after an *I-V* sweep with the highest possible current resolution and obtain off-currents of around 0.13 pA, which leads to a maximum On/Off ratio of $10^7$. The respective measurements can be found in the supporting information Figure S6. Our devices based on CMOS-compatible materials compare well with h-BN based TS devices previously reported in the literature (see supporting information Table S6). Although our transfer technique currently limits the CMOS compatibility of the fabrication process, the 2D-material community is putting substantial efforts into the development of reliable, wafer-scale transfer processes.[30,31,37] If successful, these may pave the way for an entirely CMOS-compatible process flow. Next, we performed *I-V* sweep cycling on five different devices and set different CCs for each device to investigate the device-to-device variability (as discussed for Figure 2c, the CC does not influence the $V_{th,on}$ variability). Boxplots of the extracted $V_{th,on}$ for each device and the respective CC are shown in **Figure 2d**. The corresponding single data points are plotted in the background of each boxplot. Devices D2, D4, and D5 show a rather similar $V_{th,on}$ of ~1.9 V, whereas devices D1 and D3 deviate considerably with $V_{th,on} \approx 2.45$ V. We attribute this device-to-device variability to the inhomogeneous h-BN film thickness, and we would expect a smaller variability for more homogeneous h-BN films.

An important contribution towards understanding the RS mechanism can be expected from investigating the current conduction mechanism in both HRS and LRS regimes. There are different current conduction mechanisms discussed in literature for electron transport across dielectrics, including Fowler-Nordheim tunneling, Schottky emission, Poole-Frenkel emission, and many others.[28,38] Current conduction mechanisms can be roughly classified into injection-limited, where the carrier transport depends on the metal-dielectric contact, and bulk-limited mechanisms, where the electrical transport depends on the properties of the dielectric itself.[38] Thermally activated current conduction mechanisms can be analyzed by temperature-dependent *I-V* measurements, which allows extracting physical parameters like the charge carrier mobility, trap energies, or the dielectric constant. The dominating current conduction mechanism can be determined with reasonable certainty by evaluating the goodness of the model's fit to the experimental data, although the validity of the extracted physical parameters should always be considered.

First, we analyzed the current conduction mechanism in HRS by performing temperature-dependent *I-V* measurements on a device in its HRS with a maximum applied voltage well below $V_{th,on}$. The device resistance and *I-V* sweeps were subsequently measured at different temperatures (for more details see Experimental Methods). Smoothened *I-V* curves for different temperatures between 60 K and 293 K are plotted in **Figure 3a** (the unprocessed data is provided in the supporting information section S7). The current level of the non-linear *I-V* characteristic increases with temperature (**Figure 3b**). Direct tunneling, Fowler-Nordheim tunneling, Schottky emission, and thermionic field emission are injection-limited current conduction mechanisms.[38] Fowler-Nordheim tunneling and direct tunneling have a weak temperature dependence, so that we can easily exclude them given the clear temperature dependence observed in Figure 3a. Given the fact that h-BN has a bandgap of ~6 eV[39] and that we measure at low temperatures ≤ 293 K, Schottky- and thermionic field-emission are also very unlikely in our device. From the class of bulk-limited current conduction mechanisms, we will discuss Poole-Frenkel conduction,[40–42] space-charge limited conduction (SCLC),[40,43,44] nearest neighbor hopping (NNH),[45] Mott variable range hopping (VRH),[45] hopping conduction,[45,46] and trap assisted tunneling (TAT).[47,48] We took the coefficient of determination, denoted $R^2$, to compare the goodness of the fits for the different current conduction mechanisms (see section S9 in the supporting information for more information).

An Arrhenius plot of the *I-V* curves from Figure 3a displays the natural logarithm of the current density versus the inverse temperature for four different voltages (**Figure 3c**). Two linear regimes can be identified in the Arrhenius curves, one in the temperature range from 60 K to 120 K, and the other in the temperature range from 160 K to 293 K. To visualize this, we fitted the measured data to a linear regression line in these two temperature regimes, which initially suggests the existence of two different current conduction mechanisms, each dominating in one of the respective temperature regimes. The $R^2$-values for the fits in the higher temperature range are $\geq 0.8$, indicating a reasonable level of correlation between the data and the model used. However, the $R^2$-values for the fits in the lower temperature range lie between 0.4 and 0.8, i.e. a weak correlation. In addition, the slope of the fit for 0.26 V is positive, opposite to the higher voltages. We attribute this inconclusive result to the combination of a low temperature with a low voltage, which led to an unreliable trend due to very low current levels of < 2 pA, which is close to the noise level of our measurement equipment (see Figure S7). Thus, we cannot conclusively confirm two different current conduction mechanism regimes for different temperatures.

To find the current conduction mechanism that matches best to our *I-V* curves, we linearized our data according to the current conduction mechanism theory and performed a linear regression on such data. We developed an automated algorithm to obtain the most linear region in the linearized data according to the $R^2$-criterion. A detailed explanation of the algorithm can be found in the supporting information section S8. We excluded the voltage range and corresponding values of the linearized data between 0 V and 0.26 V because of the high noise present in this range (Figure 3c). Applying the $R^2$-method to the Poole-Frenkel conduction, SCLC, and NNH current conduction mechanisms we can find parts in our *I-V* data with good linearity. However, using the extracted fit parameters to calculate the corresponding physical quantities leads to unreasonable values, so we excluded those as possible current conduction mechanisms in the HRS of our devices. To be precise, we extracted a dielectric constant of h-BN between 253 and 6281 for the Poole-Frenkel conduction, an unreasonably high concentration of free charge carriers of around $10^{17}$ cm$^{-3}$ for SCLC, and a negative hopping energy for NNH (see section S9 in the supporting information for more details). Analyzing the Mott VRH model leads to reasonable physical parameters, but the model can only describe our data for temperatures $\geq 160$ K at an applied voltage of 0.1 V, which severely limits its applicability in our case. The detailed discussion can be found in the supporting information in section S9. In the following, we discuss the hopping conduction and TAT theory in more detail, which we believe match best to our device characteristics.

In the hopping conduction theory, electrons tunnel through the insulator with the help of traps which effectively narrow the barrier width. Although the carrier energy is lower than the maximum energy of the potential barrier the carriers can still transit using the tunnel mechanism.[38] The current density depends on the temperature and the electric field as follows:[49]

$$J = qn\nu a \exp\left(\frac{qaE}{k_BT} - \frac{\Phi_t}{k_BT}\right), \qquad (1)$$

where $q$ is the elementary charge, $n$ the electron concentration in the conduction band, $\nu$ the frequency of thermal vibrations of electrons at trap sites, $E$ the applied electric field, $k_B$ the Boltzmann constant, $T$ the temperature, $a$ the mean distance between the trap sites (i.e., hopping distance), and $\Phi_t$ the energy level difference between the trap states and the conduction band minimum (see **Figure 3d**, left). The hopping distance can be extracted from the slope of the linear part in the characteristic hopping conduction plot (ln($J$) vs. $E$), shown in **Figure 3e**. The *I-V* data for each temperature is plotted with a small offset for better visibility (original data: supporting information Figure S10). The region with the gray background is the most linear region according to our $R^2$-algorithm (0.26 V – 0.76 V, $R^2$ = 0.992). The colored lines are the corresponding linear fits. The hopping distance was determined to be 2.8 ± 1.4 Å, which seems to be a reasonable (minimum) value taking note that the minimum trap distance in the current flow direction should be given by the h-BN layer spacing (~3.3 Å, see Figure 1f and supporting information section S4). Nevertheless, hopping distance values smaller than 3.3 Å could still be explained by interstitial atoms between two adjacent h-BN layers.[50] The trap energy level is extracted from the slope of the linear region in the Arrhenius plot for every applied electric field between 0.56 MV cm$^{-1}$ (0.28 V) and 1.52 MV cm$^{-1}$ (0.76 V), and we obtain a value of $\Phi_t$ = 77 ± 21 meV. The errors of the hopping distance and the trap energy are relatively large, which resembles, in a band diagram picture, the Mott VRH current conduction mechanism, where the charge carriers tunnel via traps with varying distance and energy level.[28] The analysis of the Mott VRH current conduction mechanism is based on the measurement shown in Fig. 3b and exhibit a hopping energy barrier $E_h$ (corresponding to $\Phi_t$ in the hopping conduction model) of 49 ± 8 meV at a voltage of 0.1 V (see supporting information section S9). This fits well to the extracted $\Phi_t$ in the hopping conduction theory and supports our hypothesis that the current conduction in the HRS in our devices in the low voltage regime is defect mediated.

In TAT, it is assumed that electrons tunnel through the insulator with the help of only one trap site at a certain energy $\Phi_t$ below the conduction band minimum (see Figure 3d, right)[47]. Its current density is given as[28]

$$J = A \cdot exp\left(\frac{-4\sqrt{2qm_{eff}}}{3\hbar E}\phi_t^{3/2}\right), \qquad (2)$$

where $A$ is a proportionality constant, $m_{eff}$ the electron effective mass, and $\hbar$ the reduced Planck constant. $\Phi_t$ can be extracted from the slope of the linear part in the characteristic TAT plot, shown in **Figure 3f**. The most linear part according to our R²-algorithm is between 0.68 V and 1 V (R² = 0.9985) and is indicated with a gray background in Figure 3f. The lines represent again the corresponding linear fits of the data and are plotted once more with a small offset for better visibility (original data in Figure S11). We extract a mean trap energy of $\Phi_t = 0.11 \pm 0.01$ eV by assuming $m_{eff} = 2.21 m_e$[51], where $m_e$ is the electron mass. This energy level is in good accordance with the extracted trap energy levels at smaller voltages in the hopping conduction and Mott VRH regime. In conclusion, hopping conduction and TAT current conduction mechanisms match best our temperature-dependent *I-V* data in a voltage range between 0.26 V and 1 V, while the Mott VRH current conduction mechanism can be used to describe our experimental data at 0.1 V for temperatures ≥ 160 K. All plausible current conduction mechanisms are based on defect-supported electron hopping through the h-BN. We therefore summarize that defects play a crucial role in the current conduction in the HRS. Note that not only a single trap energy level, but multiple energy levels were identified as possible trap energy levels (Hopping conduction: 77 ± 21 meV, TAT: 110 ± 10 meV). Thus, we assume that the trap energy levels responsible for the current conduction are spread over a specific energy range, which is additionally indicated by the fact that the errors in the hopping conduction theory and the hopping energy in the variable range hopping theory are relatively high.

We also performed temperature-dependent measurements on a device in a permanent LRS, which was formed by repeated *I-V* switching cycles at high CC (500 nA). This was necessary because under normal operation, the devices always fall back to the HRS once the bias voltage is tuned back towards 0 V. Temperature-dependent *I-V* curves in a temperature range between 60 K and 293 K are shown in **Figure 4a**. The current linearly depends on the voltage and the current level decreases with increasing temperature. All curves exhibit a slope very close to 1 when plotted in a double-logarithmic scale (see supporting information Figure S12), indicating ohmic conduction as the dominating current conduction mechanism.[40] The resistance vs. temperature is plotted in **Figure 4b**. The resistance linearly decreases at higher temperatures and saturates at ~360 Ω for temperatures below 80 K. The observed relation between the resistance and temperature can be explained by Matthiessen's rule, which states that the dependence of the resistivity of metals on the temperature is[52]

$$\rho(T) = \rho_0 + \rho_{ph}(T), \qquad (3)$$

where $\rho_0$ is the residual resistivity, which comprises all structural resistivity contributions like defect or surface scattering, and $\rho_{ph}(T)$ is the resistivity contribution due to electron-phonon scattering. For metals, $\rho_{ph}(T)$ depends linearly on temperature at high temperatures and the temperature dependent resistance can be described as[53]

$$R(T) = R_0 [1 + \alpha(T-T_0)], \qquad (4)$$

where $R_0$ is a certain resistance at temperature $T_0$ (often $T_0 = 0$ K), and $\alpha$ is a material-specific temperature coefficient. By fitting the temperature region $T \geq 160$ K we extracted a temperature coefficient $\alpha$ of 0.007 K$^{-1}$, which is close to the literature value of the temperature coefficient of nickel (0.006 K$^{-1}$).[54] Thus, we deduce that a stable Ni filament has been formed during repeated cycling with high CC.

For metallic filament formation, Ni ions must move through the h-BN layer under an applied voltage. Thus, the sweep rate, which is a measure of how long the voltage is applied at each measurement point, should influence $V_{th,on}$. We performed such sweep rate-dependent measurements on a TS device and plotted the results in **Figure 4c**. Although there is a visible fluctuation of $V_{th,on}$ we observe a clear trend that $V_{th,on}$ increases with increasing sweep rate, in line with our expectations. At smaller sweep rates the Ni ions have more time per voltage to move through the h-BN and reach the opposite electrode already at smaller $V_{th,on}$ values compared to higher sweep rates. **Figure 4d** (top) displays a TEM image of a device cross-section in a permanent LRS, reached after several $I$-$V$ cycles, with the bright part representing the h-BN layer. In Figure 4d (bottom), we overlay the TEM cross-sectional image with energy filtered TEM (EFTEM) images of Ni (purple) and boron (B, green). The B layer exhibits three defective regions of several nm width, highlighted by three white arrows. At these positions, the Ni electrodes are in direct contact with each other, whereas they are well separated by the h-BN layer in the rest of the image. In h-BN, the generation of B vacancies is favored compared to the formation of nitrogen (N) vacancies, which results in a higher density of N-terminated B vacancies.[50] In the EFTEM image of nitrogen (supporting information Figure S13) no distinct defects are visible in the h-BN film. This leads to the assumption that B vacancies are responsible for mediating the Ni-filament formation. The importance of B vacancies for RS of h-BN between inert electrodes was recently shown in an ab initio study of RS in two- and three-layer hexagonal boron nitride between two graphene electrodes,[55] which supports our findings.

Based on our observations and corresponding literature, we propose Ni ion diffusion through the h-BN film at B vacancies and subsequent Ni-filament formation when a positive voltage is applied to the top electrode. The facts that we observe TS behavior over three orders of magnitude of current (see Fig. 2c) and that we measure currents in the mA range after setting a device to a permanent LRS (see Fig. 4a) indicate that the filament growth is not complete when the CC limits the current. Thus, there is no direct metal contact between both electrodes and the filaments dissolve when the voltage is removed, enabling TS. The increase of the current over orders of magnitude can be explained by a decrease of the tunnel distance between the growing filaments and the counter electrode.[56] When the current through the device increases, e.g. by increasing the compliance, the filament-electrode gap decreases, and the filament needs longer time to dissolve.[57] Thus, repeated cycling at high CC can lead to a permanent LRS state with both electrodes in contact via a stable metal filament. From our TEM findings (see Fig. 4d) we deduce that not only one, but multiple filaments are formed during cycling.

## 3. Conclusion

We conducted a detailed study of the current conduction mechanisms in the high and low resistive states of threshold-switching memristors in a cross-point structure. Our memristors are based on mechanically stacked h-BN films implemented between two nickel electrodes, and thus comprise materials that are compatible with conventional silicon technology. Three different current conduction mechanisms can be discerned in the HRS: Mott variable range hopping, hopping conduction, and trap-assisted tunneling. All imply that defects play a distinct role in RS in h-BN. Characteristic ohmic conduction via Ni was found to be the dominating current conduction mechanism in the LRS. We propose Ni ion diffusion and filament formation at boron defect sites as the RS mechanism. This is consistent with the direct observation of multiple Ni filaments with TEM investigations and sweep rate-dependent *I-V* measurements. Additionally, the devices in this study show volatile operating metrics similar to state-of-the-art, including a DC-endurance of around 600 cycles with a low $V_{th,on}$ cycle-to-cycle variability of 5 %, and a high On/Off ratio of $10^7$.

## 4. Experimental Section/Methods

*Device Fabrication*: 2 cm x 2 cm Si chips covered with 90 nm thermal $SiO_2$ were used as substrates. The bottom electrodes (BE) were defined with a double layer resist stack (LOR 3A/AZ5214E from MicroChemicals GmbH/Merck Performance Materials GmbH) and optical contact lithography with an EVG 420 Mask Aligner. 60 nm of nickel (Ni) were deposited via direct current (DC) sputtering in a vonArdenne sputter tool "CS 730 S" and subsequent lift-off in a 150 C Dimethyl sulfoxide (DMSO) solution. Commercial few-layer hexagonal boron nitride (h-BN) grown on copper (Cu) foil by chemical vapor deposition (CVD) was wet-transferred with the support of a Polymethyl methacrylate (PMMA) layer. The PMMA was spin-coated onto the h-BN/Cu foil and the growth substrate was etched in an HCl + $H_2O_2$ + $H_2O$ solution. To clean the h-BN from the etchant we let the PMMA/h-BN film float on DI-water over night. After cleaning, the polymer/h-BN stack was transferred onto the $SiO_2$/Si substrate. The h-BN multilayer films thus consist of two stacked few-layer h-BN films[21]. The top electrode (TE) was defined by spin-coating of AZ5214E resist and optical contact lithography (EVG 420 Mask Aligner), sputtering of 60 nm Ni in DC-mode and subsequent lift-off in hot acetone. Contact lithography and 90 s of reactive ion etching (RIE) in an Oxford-instruments-tool "Plasmalab System 100" with a mixture of nitrogen and oxygen gas were used to remove the h-BN from the bottom contact pads. A contact lithography step, 50 nm aluminum (Al) DC-sputtering and subsequent lift-off were performed to deposit Al contact pads. See supporting information Figure S1 for a schematic of the process flow.

*Structural Device Analysis*: Optical microscope images were recorded with a Leica INM100 microscope. Raman measurements were done with a WiTec Raman spectrometer alpha300R in mapping mode with an excitation laser wavelength of 532 nm with 10 mW laser power. Scanning electron microscope (SEM) images were recorded with a Zeiss Supra 60VP SEM at an operation voltage of 4 kV. The transmission electron microscope (TEM) investigations were done in two steps: First, a thin lamella was cut out of the middle of a cross point device with a FEI Strata400 system with a gallium (Ga) ion beam and placed on a Cu grid. Second, high resolution TEM (HRTEM) and energy-filtered TEM (EFTEM) images were recorded with a Fei Tecnai $G^2$ F20 S-TWIN system at 200 kV operation voltage.

*Electrical Device Measurements*: Electrical measurements were performed in a cryogenic probe station "CRX-6.5K" from LakeShore Cryotronics connected to a semiconductor parameter analyzer (SPA) "4200A-SCS" with two source measure unit (SMU) cards "Keithley 4200-SMU", each connected to a pre-amplifier "Keithley 4200-PA" from Tektronix. The voltage was applied to the top electrode and the bottom electrode was grounded. Current-voltage (*I-V*) measurements were performed by sweeping the voltage from 0 V to a positive maximum voltage $V_{max}$ and back to 0 V. Cycling measurements were done with a 20 s delay between each sweep cycle to give the possibly formed conductive filament more time to relax[32]. The current is limited by an off-chip current limiter within the semiconductor parameter analyzer. Temperature-dependent measurements were performed in two steps, whereas the steps are repeated for each temperature: First, 15 measurement points at a constant read voltage of 100 mV were collected to extract the device resistance. In total, we measured the current for around one minute. Second, a forward sweep from -1 V to +1 V was applied. Sweep rate-dependent measurements were performed by varying the time the voltage is applied to the device before the current is measured. The set voltage ($V_{th,on}$) was automatically extracted by first smoothing the data with a Savitzky-Golay filter[58], interpolating the data to get more data points at the position of rapid current increase, and searching for the maximum in the first derivative of the smoothed, interpolated data. The reset voltage ($V_{th,off}$) is defined as the voltage where the current falls below the noise limit of the SPA, i.e., the LRS and HRS state cannot be distinguished for voltages < $V_{th,off}$. The current density $J$ was calculated by dividing the measured current by the active device area, which was measured with the optical microscope to be 34 µm². The applied electric field was calculated by dividing the applied voltage by the h-BN thickness measured with the HRTEM image (Figure 1f).

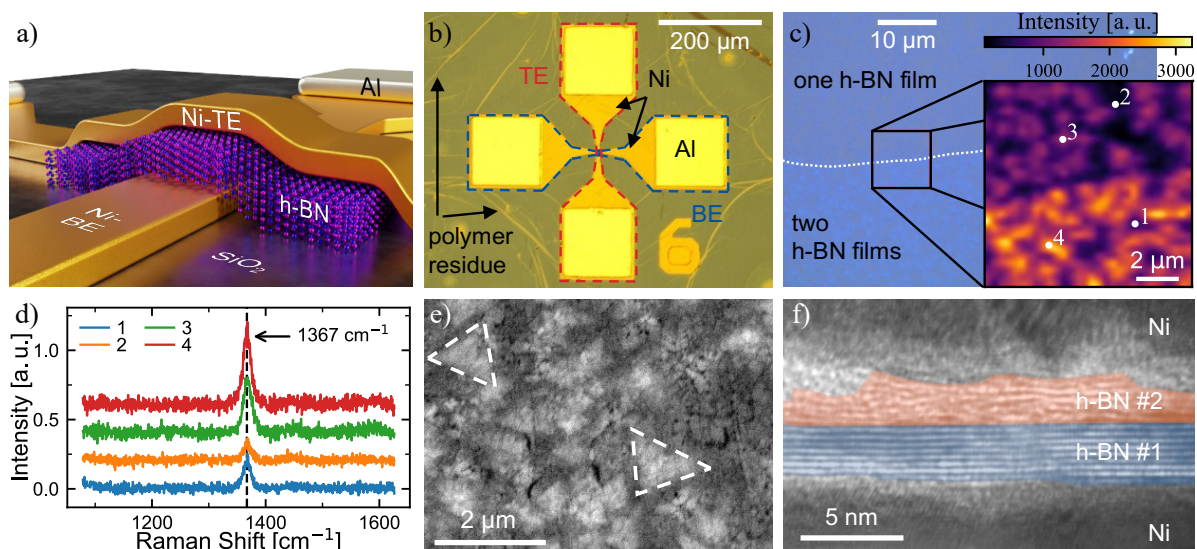

**Figure 1:** Device structure and material characterization. a) Schematic and b) optical microscope image of a Ni/h-BN/Ni cross-point device. c) Top-view optical microscope image of the as-transferred h-BN films showing the region where the first h-BN film and the stack of two h-BN films are visible. The boundary is marked by a white, dashed line. The inset shows a Raman map of the boundary region with integrated peak intensity of the characteristic h-BN $E_{2g}$ peak. d) Four different Raman spectra, extracted from the respective positions marked in c). e) Top-view SEM image of the two stacked h-BN films after transfer on $SiO_2$/Si. The white, dashed triangles highlight probable nucleation sites of multilayer grains. f) HRTEM cross-section of a Ni/h-BN/Ni cross-point device. The first and second transferred h-BN film are marked in blue and orange, respectively.

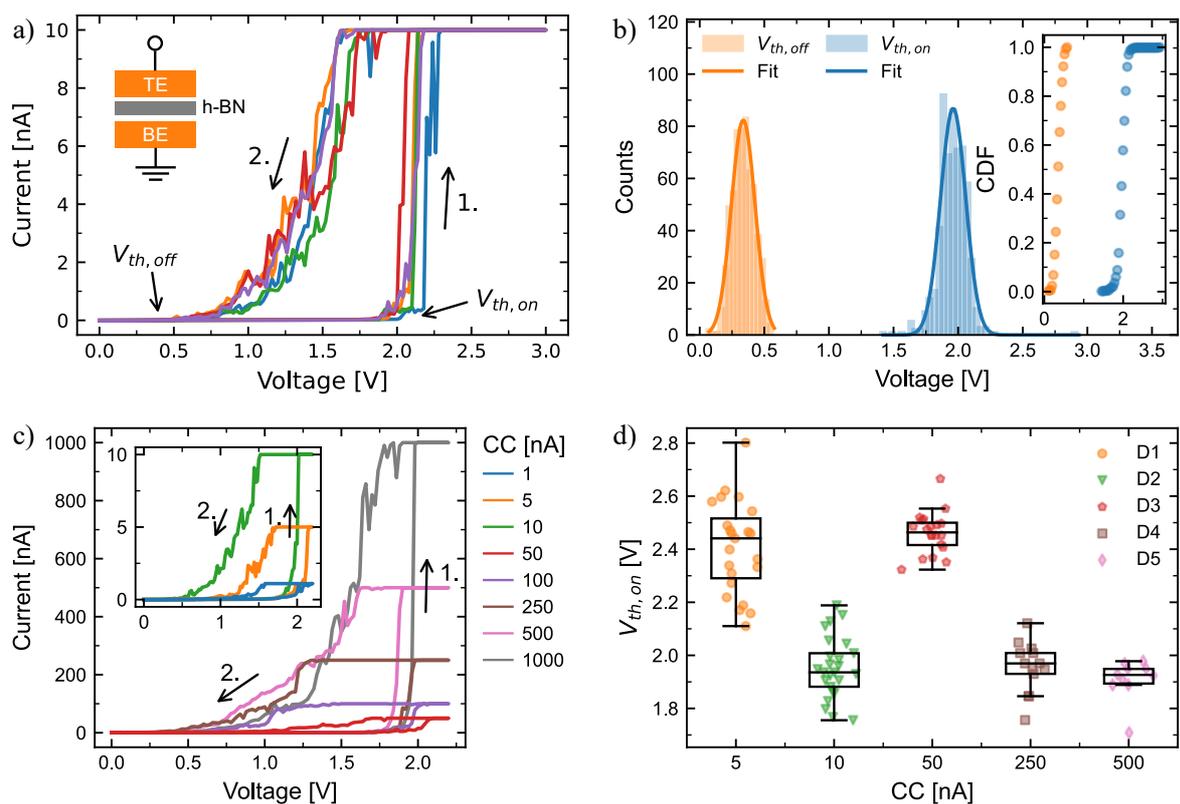

**Figure 2:** *I-V* characterization of Ni/h-BN/Ni memristors. a) Five subsequent *I-V* curves measured on a Ni/h-BN/Ni cross-point device showing TS. The arrows number one and two depict the voltage sweep direction, the others point on $V_{th,on}$ and $V_{th,off}$, respectively. The inset schematically shows the wiring during measurements. b) Histogram plot of $V_{th,on}$ and $V_{th,off}$ extracted from endurance measurement with ~600 subsequent *I-V* sweeps. The mean $V_{th,on}$ and $V_{th,off}$ of 2.0 ± 0.1 V and 0.3 ± 0.1 V, respectively, is extracted by fitting a Gaussian distribution to the histogram data. Inset: cumulative distribution function of $V_{th,on}$ and $V_{th,off}$. c) Eight subsequent *I-V* sweeps on the same device with different CCs showing TS. The CCs from 1 nA to 10 nA are plotted in the inset, the CCs from 50 nA to 1 µA are shown in the main plot. The arrows indicate the voltage sweep directions. d) Statistical analysis of $V_{th,on}$ for different CCs and different devices (D1 – D5).

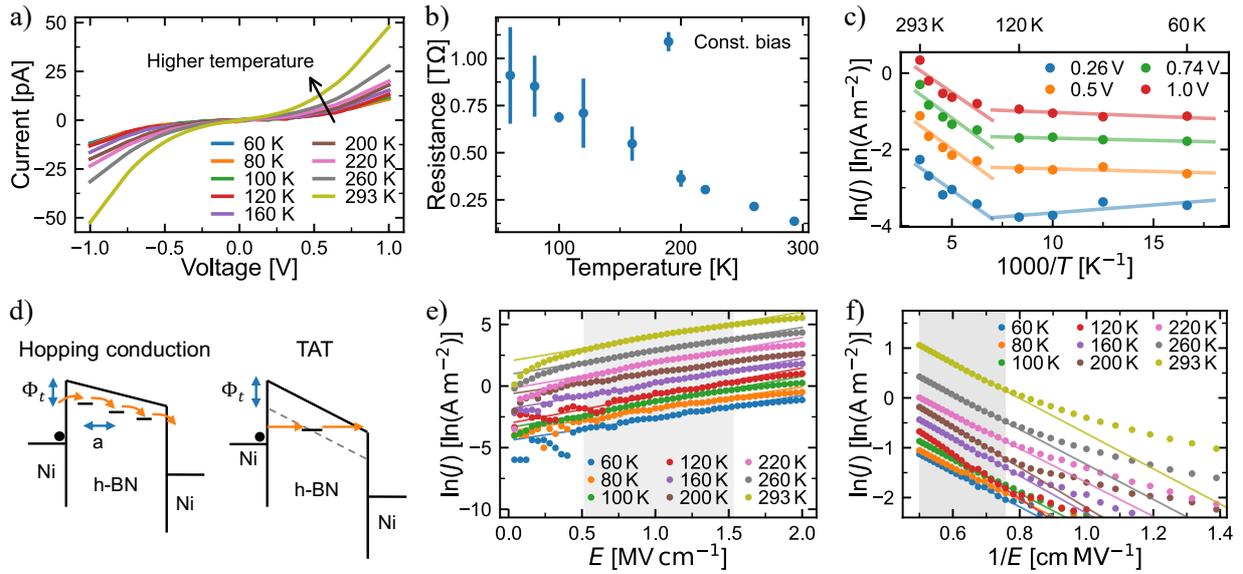

**Figure 3:** Conduction mechanism investigation of a Ni/h-BN/Ni device in HRS. a) Temperature-dependent, non-linear *I-V* curves (a Savitzky-Golay-filter was used to reduce noise, original data in Figure S7) in a temperature regime from 60 K to 293 K. b) Resistance vs. temperature measurement extracted by applying a constant voltage over time at different temperatures. c) Arrhenius plots extracted from the *I-V* data in a) showing two distinct linear regimes. The lines are linear fits to the data. d) Schematic band structure of a Ni/h-BN/Ni device for hopping conduction (left) and for TAT conduction (right). $\Phi_t$ is the trap depth in the hopping conduction and TAT equation, and $a$ is known as the hopping distance in the hopping conduction equation. e) Hopping conduction plot with an artificial offset between the curves for better visibility. The data points with a gray background (0.26 V – 0.76 V) were found to fit best with a linear model. The lines are the corresponding linear fits. f) TAT conduction plot with an artificial offset between the curves. The most linear part of the data (gray background) is found to be between 0.68 V and 1 V. The lines are the corresponding linear fits.

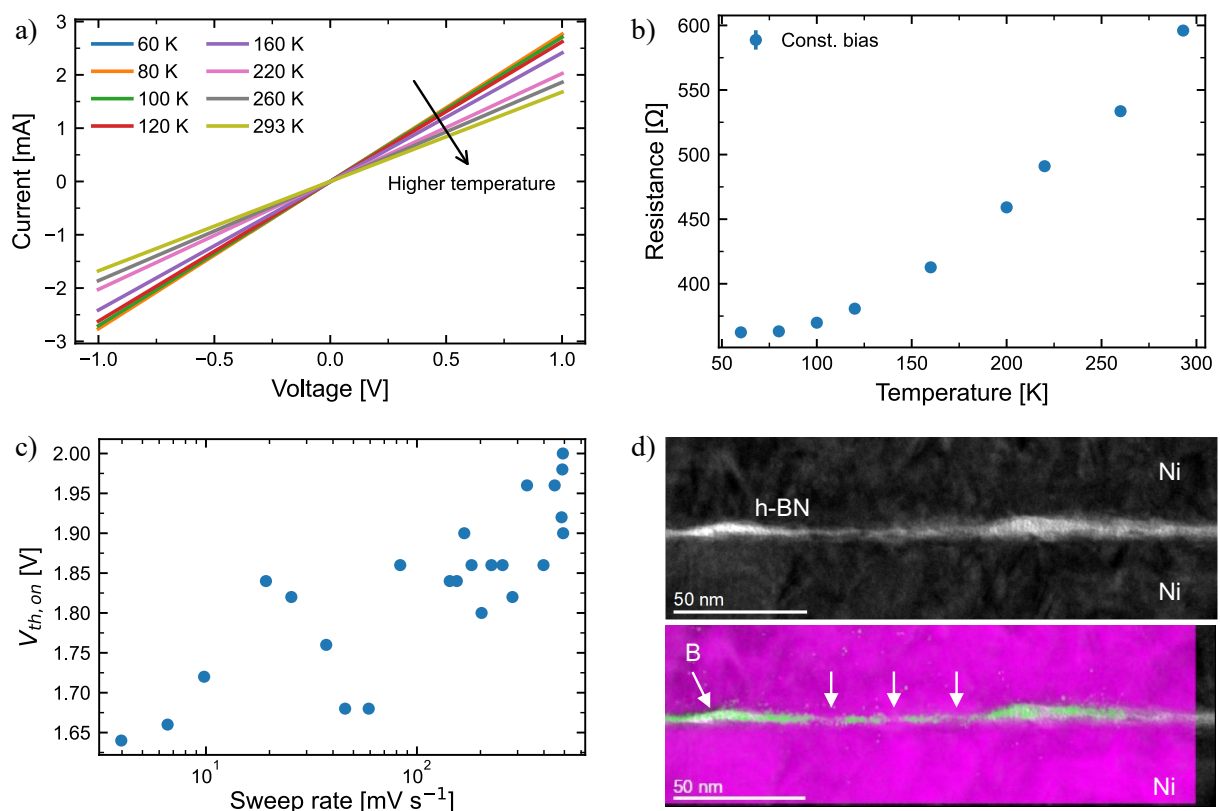

**Figure 4:** Current conduction mechanism investigation of a device in a permanent LRS state after initial switching tests. a) Temperature-dependent *I-V* curves in a temperature regime from 60 K to 293 K. b) Resistance vs. temperature measurement. c) $V_{th,on}$ vs. sweep rate of a Ni/h-BN/Ni TS device in a HRS. $V_{th,on}$ decreases with decreasing sweep rate. d) TEM image of a memristor in a permanent LRS state. Top: cross section with h-BN sandwiched between two Ni electrodes. Bottom: overlay of EFTEM images of Ni (purple) and B (green). The Ni elemental map reveals direct contact of the Ni electrodes (indicated with white arrows). Gaps in the B layer are visible at the positions where the Ni electrodes are in direct contact.

**Supporting Information**

Supporting Information is available from the Wiley Online Library or from the author.

Acknowledgements

Financial support from the German Ministry of Education and Research, BMBF, within the projects NEUROTEC (16ES1134, 16ES1133K), NEUROTEC 2 (16ME0399, 16ME0398K, 16ME0400), and NeuroSys (03ZU1106AA, 03ZU1106BB) is gratefully acknowledged.

**Conflict of Interest**

The authors declare no conflict of interest.

**Data Availability Statement**

The data that support the findings of this study are available from the corresponding author upon reasonable request.

Received: ((will be filled in by the editorial staff))
Revised: ((will be filled in by the editorial staff))
Published online: ((will be filled in by the editorial staff))

**Resistive Switching and Current Conduction Mechanisms in Hexagonal Boron Nitride Threshold Memristors with Nickel Electrodes**

*Lukas Völkel[1], Dennis Braun[1], Melkamu Belete[1], Satender Kataria[1], Thorsten Wahlbrink[2], Ke Ran[3,4], Kevin Kistermann[3], Joachim Mayer[3,4], Stephan Menzel[5], Alwin Daus[1*], Max C. Lemme[1,2]*

[1] *Chair of Electronic Devices, RWTH Aachen University, Otto-Blumenthal-Str. 25, 52074 Aachen, Germany.*

[2] *AMO GmbH, Advanced Microelectronic Center Aachen, Otto-Blumenthal-Str. 25, 52074 Aachen, Germany.*

[3] *Central Facility for Electron Microscopy (GFE), RWTH Aachen University, Ahornstr. 55, 52074, Aachen, Germany.*

[4] *Ernst Ruska-Centre for Microscopy and Spectroscopy with Electrons (ER-C 2), Forschungszentrum Jülich, 52425, Jülich, Germany.*

[5] *Peter Gruenberg Institute (PGI-7), Forschungszentrum Juelich GmbH and JARA-FIT, 52425 Juelich, Germany.*

**Section S1: Schematic process flow**

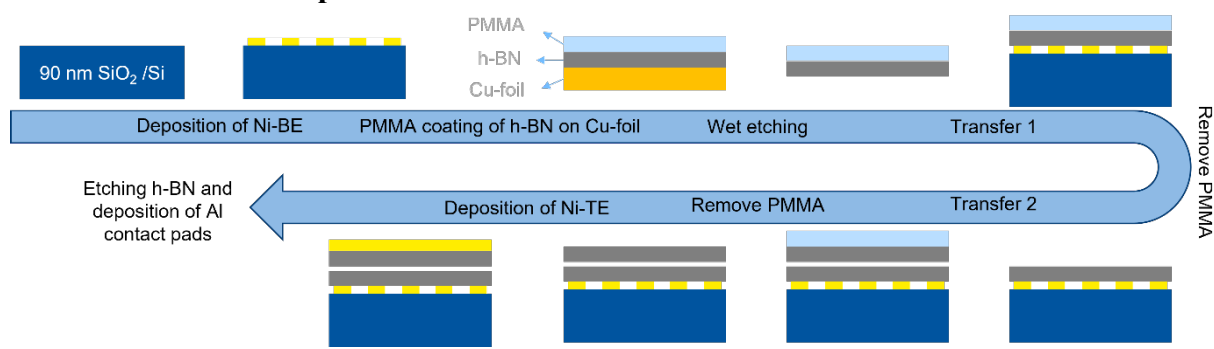

Figure S1: Schematic process flow. BE: Bottom electrode, TE: top electrode, Cu: copper, Al: aluminum, Ni: nickel.

**Section S2: Current-voltage measurements on devices without h-BN**

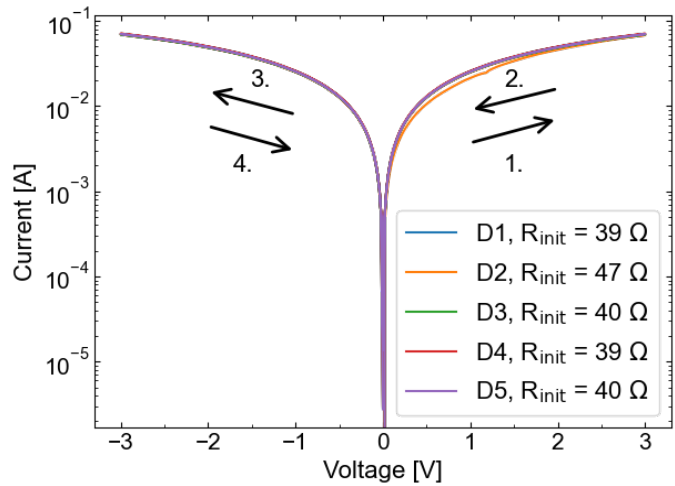

Figure S2: *I-V* curves measured of five different Ni/Ni memristors D1-D5.

To measure the possible influence of a nickel oxide layer between the Ni electrodes or between Ni and the Al contact pads on the RS behavior, we measure the initial resistance of devices without h-BN between the electrodes and subsequently apply a positive and a negative sweep with a maximum voltage of ±3 V. The initial resistance of all devices is only a few Ω and none of the devices show resistive switching. Thus, we can exclude an influence of the electrode material system on our resistive switching measurements.

**Section S3: Histogram of h-BN $E_{2g}$ peak position from Raman map spectra**

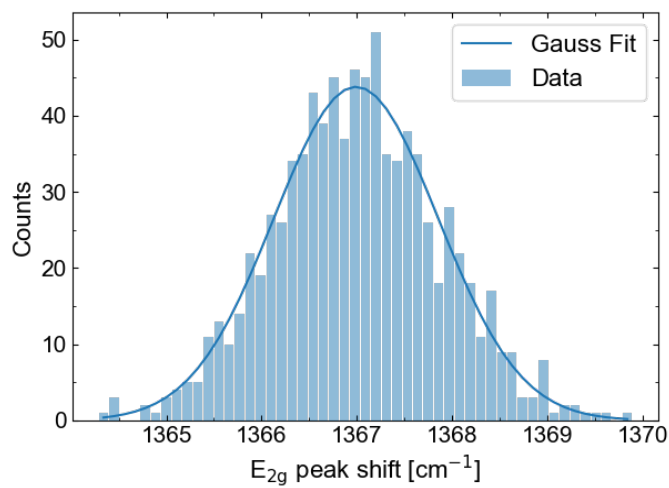

Figure S3: Histogram of extracted $E_{2g}$-peak position of all spectra in the Raman map. The peak positions are extracted by fitting a Lorentz-Peak to each Raman spectrum. A Gaussian distribution function with the center at 1367.0 cm$^{-1}$ and a distribution width of $\sigma$ = 0.9 cm$^{-1}$ is fitted to the histogram data, which corresponds to 3-layer to bulk h-BN[1].

**Section S4: HRTEM and van der Waals spacing**

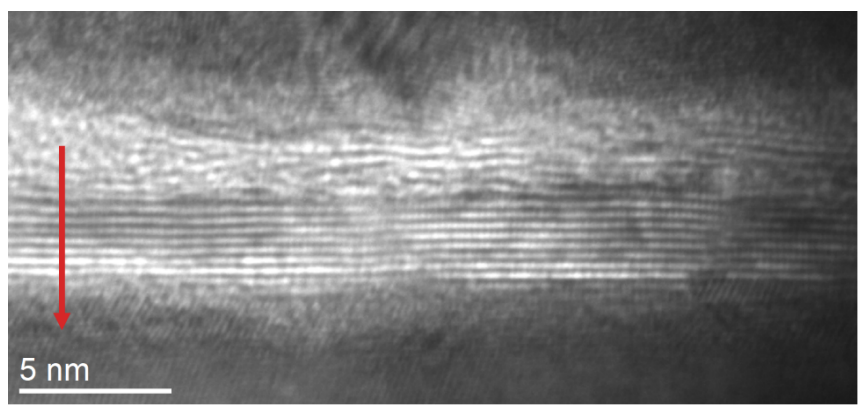

Figure S4-1: HRTEM image with the arrow indicating the position for line profile extraction.

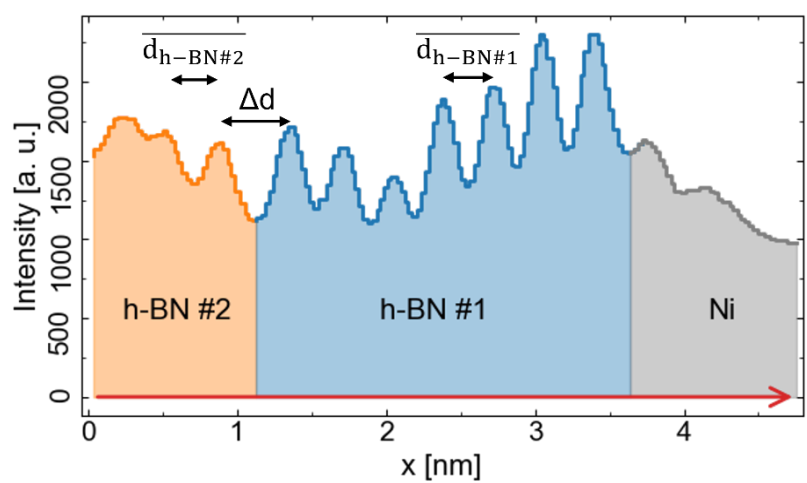

Figure S4-2: From HRTEM extracted line-profile. The orange, blue and gray regions are identified as h-BN film #2, h-BN film #1, and the Ni bottom electrode, respectively. $\overline{d_{h-BN\#1}} = 0.321 \pm 0.029$ nm is the mean interlayer distance of two adjacent h-BN layers in h-BN film 1, $\overline{d_{h-BN\#2}} = 0.340 \pm 0.016$ nm is the mean interlayer distance of two adjacent h-BN layers in film 2, and $\Delta d = 0.479$ nm is the spacing between h-BN film 1 and 2. The measured interlayer distance agrees with other literature values[2,3], whereas $\Delta d$ is almost 30 % bigger.

**Section S5: *I-V* Curves of endurance measurement**

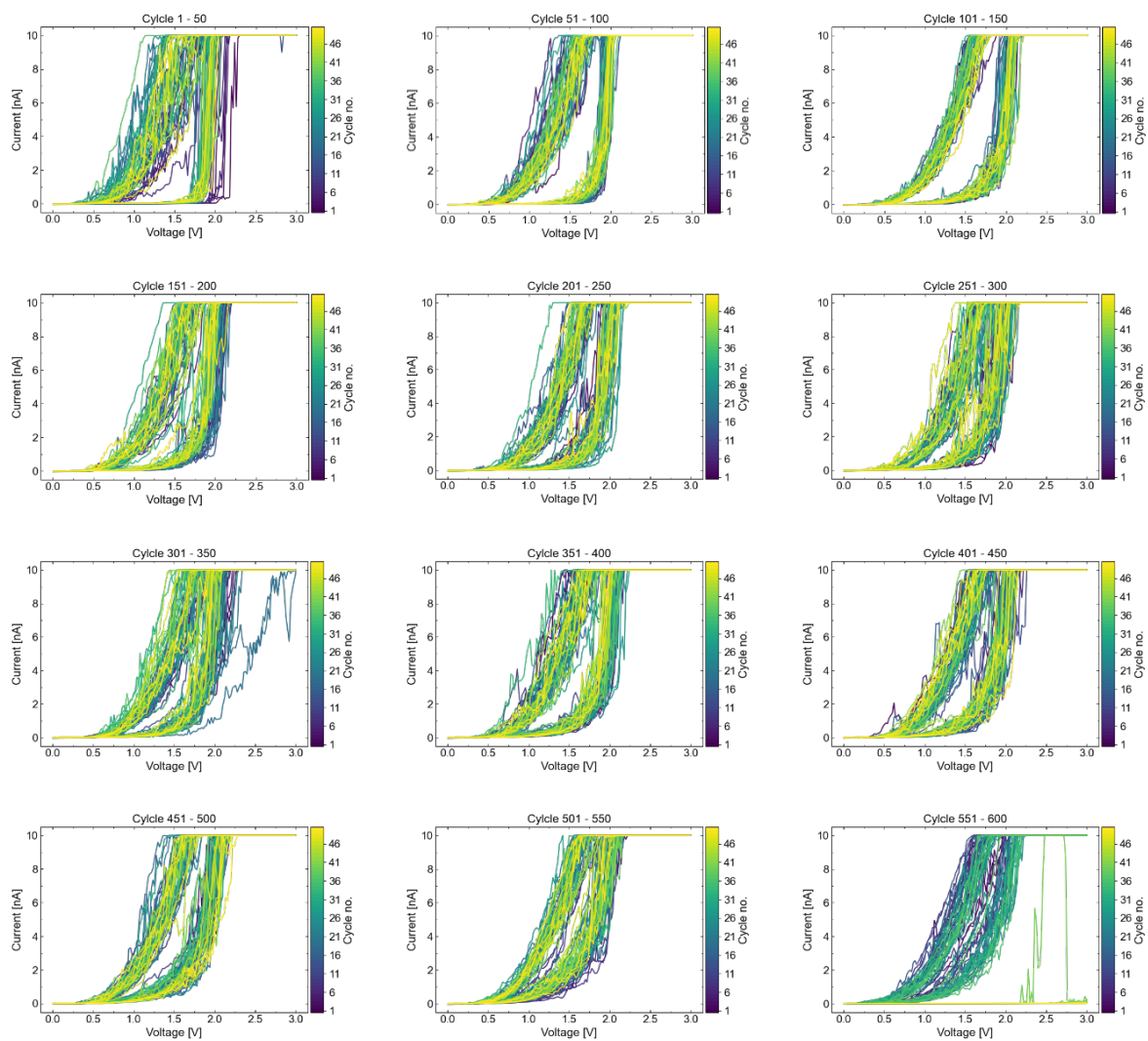

Figure S5: Complete set of *I-V* cycles measured during the endurance measurement. Every plot contains 50 cycles. The title of the plots denotes the respective cycle number range.

**Section S6: Comparison of device performance with other literature**

| $I_{cc,max}/I_{cc,min}$ [A] | max. On/Off ratio | max. DC Endurance | Electrode material (BE/TE) | h-BN thickness | CMOS-compatible configuration | Reference |
|---|---|---|---|---|---|---|
| $10^2$ | $5 \cdot 10^4$ | 5 | Cu/Pt | 3.5 nm | No (C-AFM) | [4] |
| 10 | $5 \cdot 10^5$ | 7 | Pt/Ag | 1.5 nm | No | [5] |
| 5 | $5 \cdot 10^7$ | - | Ag/Pt, Ag/Ag | 0.33 nm | No, No | [6] |
| $10^2$ | $10^4$ | 2 | Ti/Cu | 5-7/15-20 layers | Yes | [7] |
| 0 | $10^5$ | 4 | Fe/Ag | ~ 15 layers | No | [8] |
| 0 | $10^3$ | - | Fe/Ag | > 10 layers | No | [9] |
| 0 | $5 \cdot 10^6$ | ~ few 10 | Ag/Au | 9-15 layers | No | [10] |
| $10^9$ | $10^{11}$ | - | Ag/Ag | 1.5 nm | No | [11] |
| 0 | $5 \cdot 10^5$ | ~ several 10 | Ag/Pt, Pt/Pt | 1.5 nm | No, Yes | [11] |
| $10^2$ | $5 \cdot 10^4$ | ~ 300 | Ti/Cu | 1.8 nm | Yes | [12] |
| 0 | $10^5$ | ~ few 10 | Ag/Au | 15-18 layers | No | [13] |
| 0 | $10^4$ | - | Au/Ni, Ni/Ni, Pt/Ni | 1 layer | No (common BE) | [14] |
| $10^5$ | $10^8$ | 100 | Pt/Ag, Pt/Gr/Ag | 1 layer | No | [15] |
| **$10^3$** | **$10^7$** | **588** | **Ni/Ni** | **5 ± 1 nm** | **Yes** | **This work** |

Table S6: Benchmarking table of publications showing TRS in h-BN-based resistive switching devices. $I_{cc,max}/I_{cc,min}$ denote the current compliance window in which the device shows threshold resistive switching behavior. A CMOS-compatible configuration includes a CMOS compatible material stack as well as an application-near device structure like cross point devices. Devices with a common bottom electrode or conductive atomic force microscopy (C-AFM) investigations are considered as not CMOS compatible configurations.

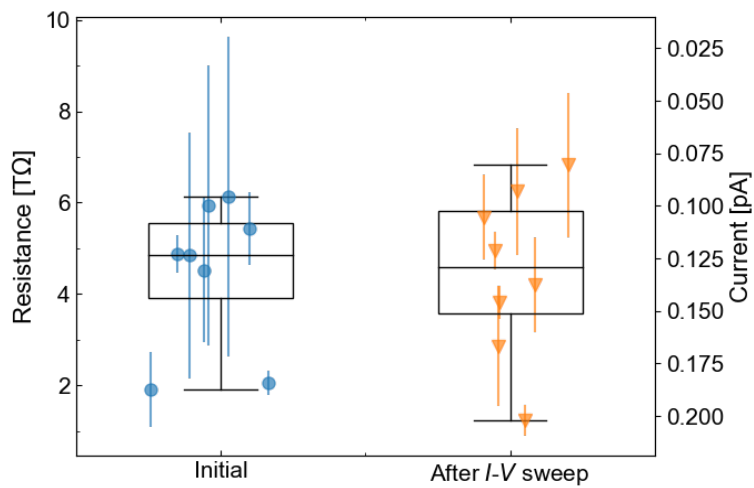

Figure S6: Boxplots of resistances and current levels for 8 different devices. "Initial" represent measurements before any volatile RS events and "After *I-V* sweep" is after performing a RS measurement showing that resistance levels return to the original state. The single measurement points are plotted in the background of each boxplot. The resistance was measured at a voltage of 0.1 V. 15 measurement points were recorded per measurement. The measured current level after the *I-V* sweep serves as the off current for the calculation of the maximum On/Off ratio in table S6.

**Section S7: Temperature dependent *I-V* curves of Device in HRS**

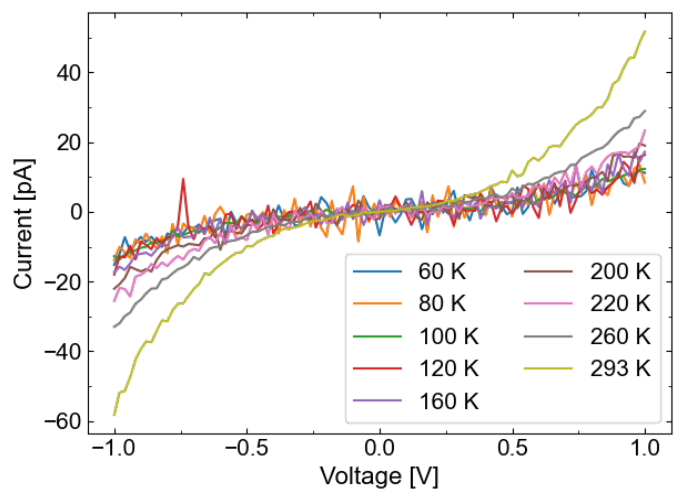

Figure S7: Unfiltered *I-V* data of the temperature dependent measurements of a Ni/h-BN/Ni device in HRS. The *I-V* curves shown in the main paper are smoothened with a Savitzky-Golay filter[16].

**Section S8: Algorithm to get the most linear region of a curve according to R²-criterion**

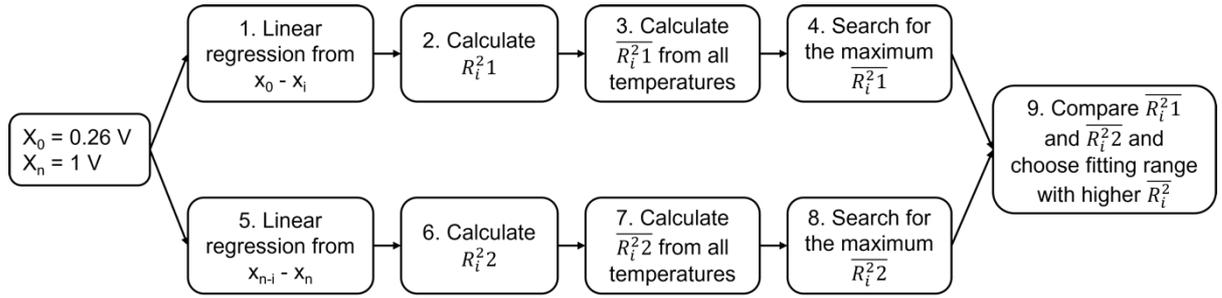

Figure S8: Flow-chart of algorithm to get the most linear region in our *I-V* curve.

The R²-value is defined as the ratio of the squared deviation of the fitted values from the measured values and the squared deviation of the measured values from the global mean of a linear regression model, and normalized to an absolute value between 0 and 1[17,18]. A value close to 1 denotes a good match of the linear fit to the data, whereas a value close to 0 denotes a bad match of the linear fit to the data.

We exclude the voltage range between 0 V and 0.26 V from our fits because in Figure 3c in the main paper one can see that at a voltage of 0.26 V the measurement noise is high, which will lead to inconsistent results. This means for the fitting algorithm that the minimum fitting range is fixed to $x_0$ = 0.26 V or its corresponding x-value of the linearized *I-V* data corresponding to the conduction mechanism formula. The following steps are performed to find the most linear region in our curves according to the R²-criterion with the constraint that either the maximum or the minimum of the fitting range is fixed:

1. Perform a linear regression from $x_0$ to $x_i$ for each temperature, where $x_i$ is the i-th data point and its range is $i = \{10, 11, 12, \dots, n-1, n\}$, where $n$ is the total number of data points.
2. Calculate $R_i^2 1$ for each linear regression. With $i \geq 10$ the smallest linear regression includes 10 data points.
3. Calculate $\overline{R_i^2 1}$, the mean $R_i^2 1$ from all temperatures.
4. Search for the maximum $\overline{R_i^2 1}$.
5. Perform a linear regression from $x_{n-i}$ to $x_n$, with $i = \{10, 11, 12, \dots, n-1, n\}$.
6. Calculate $R_i^2 2$ for each linear regression.
7. Calculate $\overline{R_i^2 2}$, the mean $R_i^2 2$ from all temperatures.
8. Search for the maximum $\overline{R_i^2 2}$.
9. Compare $\overline{R_i^2 1}$ and $\overline{R_i^2 2}$ and choose the fitting range with the higher $\overline{R_i^2}$.

The fitting range with the highest $\overline{R_i^2}$ corresponds, according to the R²-criterion, to the most linear region in the curves. The algorithm is summarized in Figure S8.

**Section S9: Discussion of various current conduction mechanisms**

In this section we discuss the applicability of the Poole Frenkel model, the space charge limited conduction (SCLC) model, the nearest neighbor hopping (NNH) and the Mott variable range hopping (VRH) model to our temperature dependent *I-V* data in the HRS.

The Poole Frenkel CCM assumes that electrons are thermally emitted from traps into the conduction band of the dielectric material and it has a characteristic electric field and temperature dependence of[19]

$$J \sim E \cdot [\frac{-q\left(\Phi_T - \sqrt{\frac{qE}{\pi \epsilon_r \epsilon_0}}\right)}{k_B T}], \tag{s1}$$

where $E$ is the electric field, $q$ is the elementary charge, $\Phi_t$ is the trap depth, $\epsilon_r$ is the dielectric constant, $\epsilon_0$ is the vacuum permittivity, and $k_B$ is the Boltzmann constant. Considering equation (s1), a plot of $\ln(J/E)$ vs. $\sqrt{E}$ should be linear. The electric field is calculated by dividing the applied voltage by the h-BN film thickness extracted from the HRTEM image in Figure 1f in the main paper. The Poole-Frenkel plot is shown in Figure S9-1. We found the most linear region in the Poole-Frenkel plot between 0.56 V and 1 V. According to equation (s1), we can calculate the dielectric constant of our h-BN film with the slope extracted by the linear regression. The calculated dielectric constants for all temperatures can be found in Table S9-1. We see that the dielectric constant highly depends on the temperature and additionally has values more than 50 times higher than the literature value of $\epsilon_r = 5$[20]. Thus, we exclude the Poole Frenkel CCM as the responsible CCM in our devices.

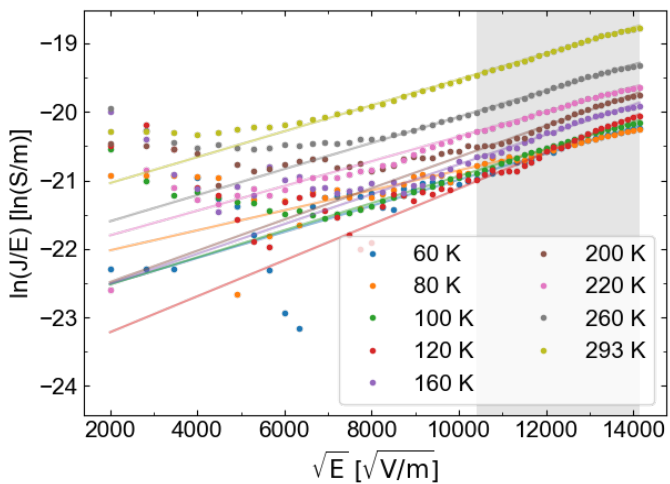

Figure S9-1: Poole-Frenkel plot for all temperatures. The data points with a gray background (10392 $\sqrt{V/m}$ - 14142 $\sqrt{V/m}$, which corresponds to 0.54 V – 1 V) fit best to a linear model. The lines are the corresponding linear fits.

| Temperature [K] | Dielectric constant $\epsilon_r$ |
|---|---|
| 60 | 6281 |
| 80 | 5644 |
| 100 | 2047 |
| 120 | 769 |
| 160 | 626 |
| 200 | 357 |
| 220 | 489 |
| 260 | 319 |
| **293** | 253 |

Table S9-1: Extracted dielectric constants for all temperatures.

The SCLC theory predicts the current density between two plane, parallel electrodes to be split in three regimes, the ohmic conduction current ($J_{Ohm} \sim V$), the trap-filled limit current ($J_{TFL} \sim V^2$) and Child's law ($J_{Child} \sim V^2$), where the current densities depend on the applied voltage as follows:[21]

$$J_{Ohm} = qn_0\mu\frac{V}{d}, \qquad (s2)$$

$$J_{TFL} = \frac{9}{8}\mu\epsilon_r\epsilon_0\Theta_0\frac{V^2}{d^3}, \qquad (s3)$$

$$J_{Child} = \frac{9}{8}\mu\epsilon_r\epsilon_0\frac{V^2}{d^3}. \qquad (s4)$$

$n_0$ is the concentration of the free charge carriers in thermal equilibrium, $\mu$ is the charge carrier mobility, $\Theta_0$ is the ratio of the effective density of states in the conduction band and the density of traps inside the h-BN, d is the distance between the electrodes, and $V$ is the applied voltage. Between $J_{TFL}$ and $J_{Child}$ is a transition region where $J \sim V^x$ with $x > 2$. We expect to be only in the first two regions, as we apply relatively low voltages. For this CCM we do not apply our algorithm for maximized R²-value but adapt it to find the fitting range leading to a slope of 2 (with a fixed maximum of the fitting range at 1 V) and the fitting range leading to a slope of 1 (with a fixed minimum of the fitting range at the second data point at 0.06 V). This procedure leads to an overlap of the fitting ranges, so we adjust the overlap region manually in a way that there is no overlap and at the same time the R² values remain close to 1. In Table 9-2 the extracted slopes of the linear regression fits can be found. With the equations (s2) and (s3) one can extract the mobility of the charge carriers from the intercept of the fitted lines. However, in the case of the ohmic conduction region, we must estimate $n_0$ to calculate the mobility, and in the case of the TFL region, we must estimate $\theta$. We will only extract the mean mobility from all temperatures as we must estimate the values for $\theta$ and $n_0$ anyways, which is a relatively big error source. A. Rose[22] gives an approximate value for $\theta$ of $\sim 10^{-7}$. With this, we estimate the mean mobility in the TFL region of all temperatures to be $\mu_{mean} = (1.42 \pm 0.88) \cdot 10^{-4} \frac{cm^2}{Vs}$, which is comparable to the values extracted by Chiu et. al.[23,24] for other insulators. However, the extraction of a realistic value for the mobility in the ohmic region is not possible. We found no applicable literature value for $n_0$. To get a first estimation of the mobility we set $n_0$ to 1. This results in a mean mobility of $\mu_{mean} \approx 5 \cdot 10^{13} \frac{cm^2}{Vs}$. This means, $n_0$ must be in the order of $10^{17} \frac{1}{cm^3}$ to get a comparable value for the mobility in both regions, which is unrealistic for a wide-bandgap, insulating material like h-BN at low temperatures (< 260 K).

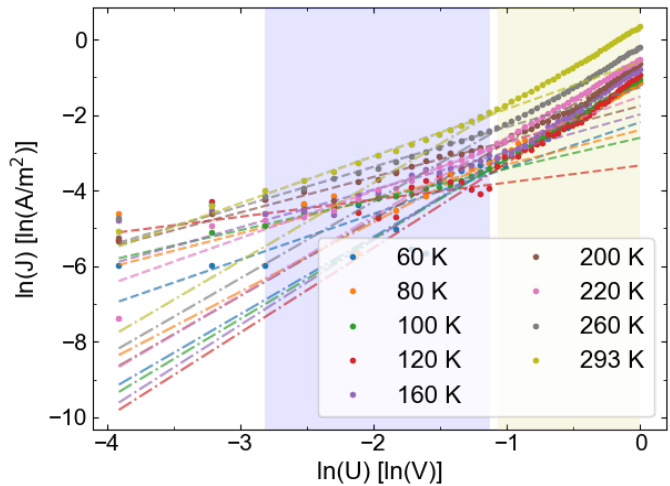

Figure S9-2: Characteristic space charge limited CCM plot for all temperatures. The blue and yellow regions mark the fitting range for ohmic conduction regime and trap-filled limit current regime, respectively. The dashed and dash-dotted lines are the fits of the blue and yellow region, respectively.

| T [K] | Slope blue region | R² blue region | Slope yellow region | R² yellow region |
|---|---|---|---|---|
| 60 | 1.21 | 0.448 | 2.05 | 0.993 |
| 80 | 0.91 | 0.526 | 1.84 | 0.994 |
| 100 | 0.81 | 0.975 | 2.12 | 0.999 |
| 120 | 0.45 | 0.368 | 2.24 | 0.985 |
| 160 | 1.00 | 0.916 | 2.25 | 0.996 |
| 200 | 0.94 | 0.940 | 2.03 | 0.987 |
| 220 | 1.24 | 0.979 | 2.08 | 0.999 |
| 260 | 1.05 | 0.995 | 2.03 | 0.998 |
| 293 | 1.25 | 0.994 | 2.06 | 0.999 |

Table 9-2: Extracted slopes of linear regression fits for different temperatures in respective regions of the characteristic SCLC plot.

The NNH and Mott VRH mechanisms assume electrons hopping through the dielectric from traps to traps via nearest neighboring traps with a certain barrier height, or traps with a variable (wider) distance but possibly a lower energy barrier, respectively. The models describe a characteristic temperature dependence of[25]

$$\sigma = \sigma_0 \cdot e^{-(T_0/T)^\beta}, \quad (s5)$$

where $\sigma$ is the conductance, $\sigma_0$ is the conductivity at a certain temperature $T_0$, $T$ is the temperature, and $\beta$ is a parameter dependent on the model under consideration. In the NNH and Mott VRH models $\beta = 1$ and $\beta = \frac{1}{4}$, respectively. In contrast to the CCMs discussed before we need to measure the temperature dependent conductance of our devices to apply these models. This is done by calculating the inverse of the resistance shown in Figure 3b in the main paper, which was measured at a voltage of 0.1 V. By extracting $T_0$ one can calculate the hopping energy[26]

$$E_h(T) = \frac{1}{4} k_b T^{3/4} T_0^{1/4}, \quad (s6)$$

which reflects the energy barrier between defects. From equation (s5), one expects a linear behavior when the natural logarithm of the conductance is plotted over $\frac{1}{T^\beta}$. The characteristic plots for NNH and Mott VRH are shown in Figures S9-3 and S9-4, respectively. Both plots exhibit a nonlinear behavior over the full temperature range, but a linear region for temperatures $\geq 160$ K can be identified in both plots. In the smaller temperature range the data points seem to be linear, too. However, having a closer look to the fit statistics let us exclude these temperatures from our further analysis due to the high noise in the measurement (compare argumentation to the Arrhenius plot discussed in the main paper). According to equation (s5), $T_0$ depends on the slope of the linear fit and $\sigma_0$ is related to the intercept of the linear fit. Conducting a linear fit for temperatures $\geq 160$ K in the NNH plot we get $T_0$ = -564 K and $\sigma_0$ = 44 pS. The negative $T_0$ would lead to a negative hopping energy, which is physically not meaningful. Thus, we exclude NNH as a possible conduction mechanism.

We find $T_0$ = 23.31·10$^5$ K and $\sigma_0$ = 84 nS by fitting the region for $T \geq 120$ K in the Mott VRH characteristic plot. According to equation (s6), we get a mean $E_h$ = 49 ± 8 meV.

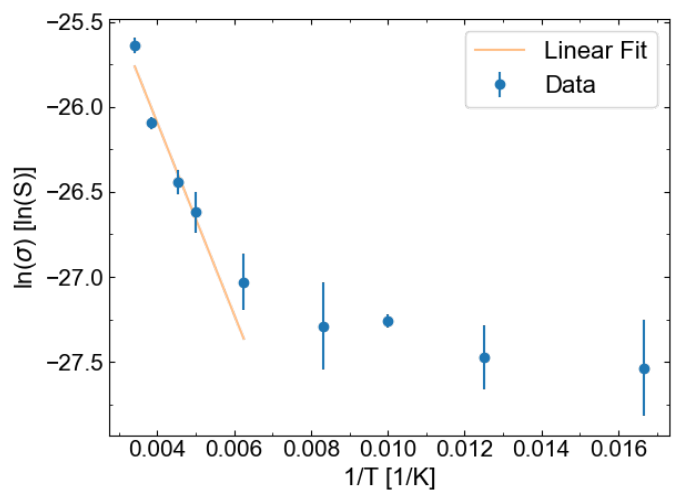

Figure S9-3: Nearest neighbor hopping plot.

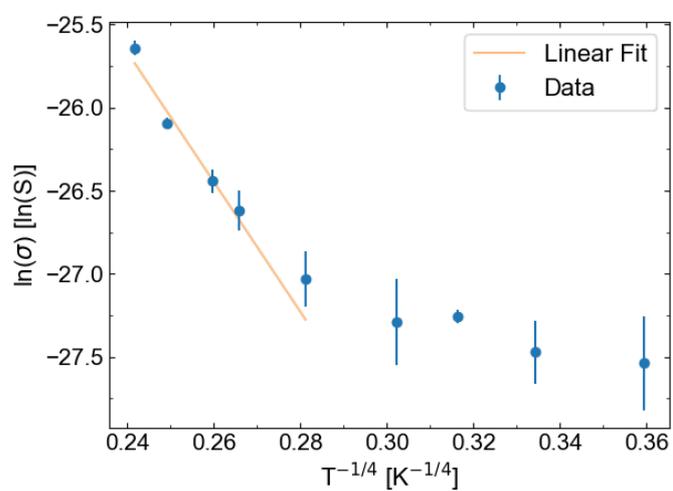

Figure S9-4: Mott variable range hopping plot.

**Section S10: Hopping conduction**

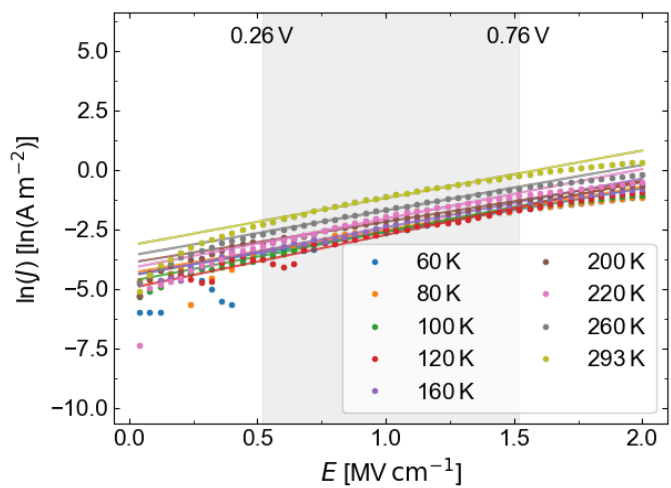

Figure S10: Characteristic hopping conduction plot without an additional offset.

**Section S11: Trap assisted tunneling**

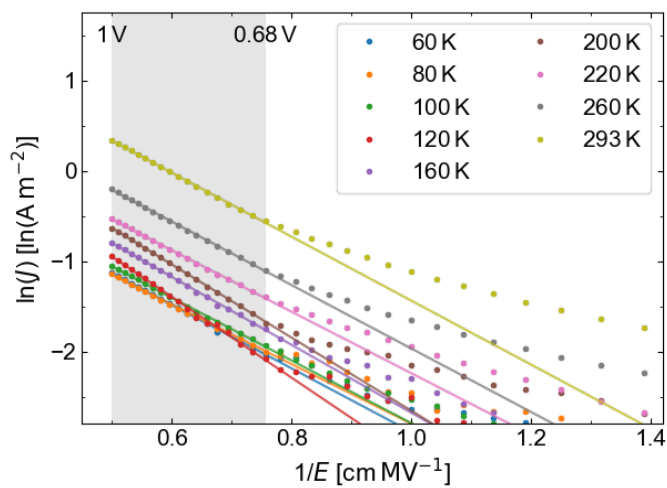

Figure S11: Characteristic trap assisted tunneling plot without an additional offset.

**Section S12: Temperature dependent I-V measurements of a device in permanent LRS**

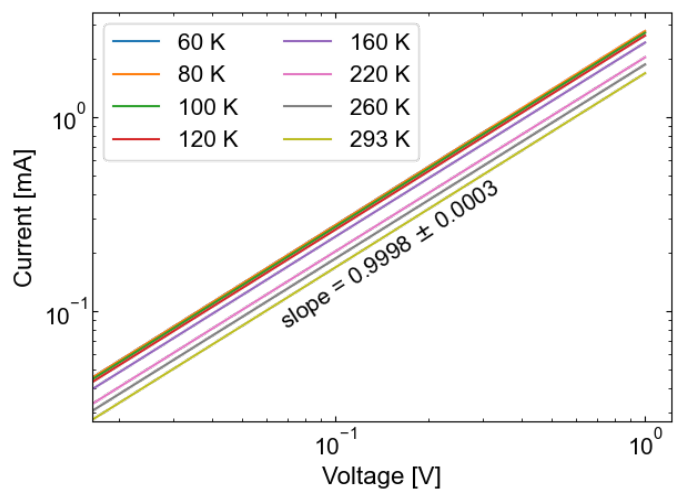

Figure S12: Double-logarithmic plot of I-V curves shown in Figure 4a in the main paper. All curves show clear linear behavior with a mean slope of 0.9998 ± 0.0003 A/V.

**Section S13: Energy filtered TEM (EFTEM) images**

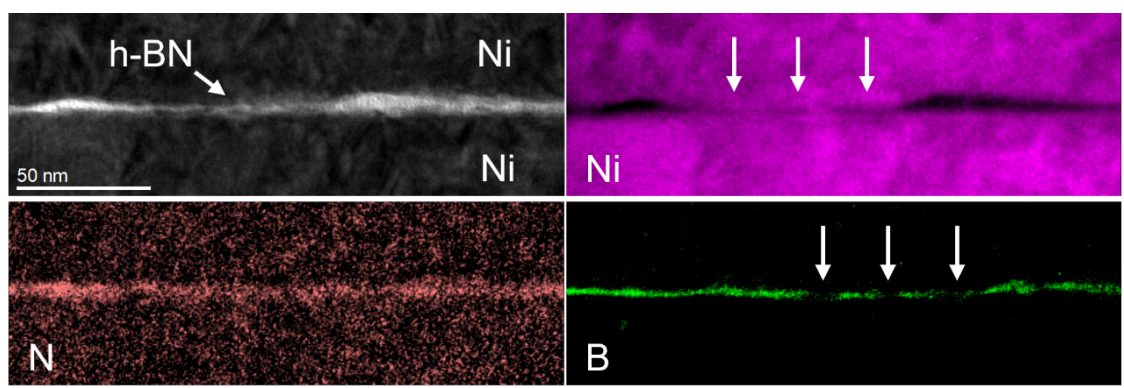

Figure S13: Cross-section TEM and EFTEM images of nickel, nitrogen, and boron. Due to drift of the TEM-lamella between the different EFTEM measurements the marked features do not appear at the same positions. However, referencing the EFTEM maps to the cross-section TEM image let us create the overlay-image shown in Figure 4d in the main paper. The nitrogen map reveals no distinct defects in the h-BN film.